\documentclass[a4paper]{article}
\usepackage{geometry}
\usepackage{graphicx}
\usepackage{amssymb}
\usepackage{graphicx}
\usepackage{braket}
\usepackage{cases}
\usepackage{algorithm}
\usepackage{algorithmic}
\usepackage{tikz}
\usepackage{makecell}
\usepackage{booktabs}
\usepackage{subfigure}
\usepackage{indentfirst}
\usepackage{hyperref}
\usepackage{siunitx}
\title {Solving Multi-Dimensional Stationary Schr\"{o}dinger Equations Using Extended Physics-Informed Neural Networks}
\author{Jinde Liu, Xilong Dou, Chen Yang\thanks{yangchen@scu.edu.cn}, Gang Jiang}
\date{\today}
\begin{document}

\maketitle

\begin{abstract}
    Due to the good performance of neural networks in high-dimensional and nonlinear problems, machine learning is replacing traditional methods and becoming a better approach for eigenvalue and wave function solutions of multi-dimensional Schr\"{o}dinger equations. This paper proposes a numerical method based on neural networks to solve multiple excited states of multi-dimensional stationary Schr\"{o}dinger equation. We introduce the orthogonal normalization condition into the loss function, use the frequency principle of neural networks to automatically obtain multiple excited state eigenfunctions and eigenvalues of the equation from low to high energy levels, and propose a degenerate level processing method. The use of equation residuals and energy uncertainty makes the error of each energy level converge to 0, which effectively avoids the order of magnitude interference of error convergence, improves the accuracy of wave functions, and improves the accuracy of eigenvalues as well. Comparing our results to the previous work, the accuracy of the harmonic oscillator problem is at least an order of magnitude higher with fewer training epochs. We complete numerical experiments on typical analytically solvable Schr\"{o}dinger equations, e.g., harmonic oscillators and hydrogen-like atoms, and propose calculation and evaluation methods for each physical quantity, which prove the effectiveness of our method on eigenvalue problems. Our successful solution of the excited states of the hydrogen atom problem provides a potential idea for solving the stationary Schr\"{o}dinger equation for multi-electron atomic molecules.
    

\end{abstract}

\section{Introduction}

Solving the eigenvalue problem of the multi-dimensional stationary Schr\"{o}dinger equation (SSE) has been an issue since quantum mechanics established that many physicists pay attention to. The solution to this problem is of great significance to the development of quantum chemistry and computational materials science. Except for a few models that can be solved analytically, e.g., infinitely deep potential wells, harmonic oscillators, the hydrogen atom, Morse potentials, etc., the SSE can only offer approximate solutions.

As a specific eigenvalue problem, the solution of the SSE has great versatility with other types of eigenvalue problems. The key process of finding an approximate solution consists of the representation method of the eigen wave function, the model transformation, and the solving procedure. The model transformation and the solving procedure are generally needed to transform the equation-solving problem into an optimization problem, and then use the optimization algorithm to solve it. More generally, for the eigenvalue problem of partial differential equations (PDE), the solving process has the following key points: problem transformation, eigenvalue solving, normalization condition (non-zero solution construct), orthogonal condition (excited state treatment), boundary conditions, and sampling method. In view of the above key points, different numerical calculation methods based on neural networks are proposed \cite{noauthor_artificial_1997,nakanishi_numerical_2000,noauthor_numerical_2001,shirvany_numerical_2008,e_deep_2018,han_solving_2020,ben-shaul_solving_2020,li_neural-network-based_2021,zhang_solving_2022,jin_physics-informed_2022,yang_neural_2022,li_semigroup_2022}.  

Lagaris \emph{et al.}\cite{noauthor_artificial_1997} proved for the first time the feasibility of using neural networks to solve eigenvalue problems. Nakanishi \emph{et al.}\cite{nakanishi_numerical_2000,noauthor_numerical_2001} was trained to solve the eigenvalues as parameters for the first time, and the excited state at high energy levels was calculated. Shirvany \emph{et al.}\cite{shirvany_numerical_2008} adds boundary conditions to the loss function to obtain the solution of multiple energy levels of an one-dimensional harmonic oscillator. After the extensive development of neural networks, E \emph{et al.}\cite{e_deep_2018} for the first time used the variational form of the residual network and eigenvalue problem to solve the ground state of the high-dimensional Schr\"{o}dinger equation. Han \emph{et al.}\cite{han_solving_2020} for the first time used the backward stochastic differential equations form of diffusion Monte Carlo to solve the ground and excited states of the high-dimensional Schr\"{o}dinger equation in combination with the neural network, and the automatic constraint of periodic boundary conditions is realized by adding a layer of trigonometric functions to the network. Li \emph{et al.}\cite{li_neural-network-based_2021} constructs a multi-output network to represent multiple eigenstates, solves the multi-dimensional harmonic oscillator problem, proposes an algorithm that obtains multiple states in a single training, and discusses the resource consumption problem when solving high-dimensional problems. Zhang \emph{et al.}\cite{zhang_solving_2022,jin_physics-informed_2022,yang_neural_2022,li_semigroup_2022,quantum_wells} also made some explorations on the issue of eigenvalues.

There are some limitations of previous methods, e.g., only low-dimensional cases and a small number of excited states were solved, degeneracy problems were rarely discussed, and the solution accuracy obtained was poor. At present, there is still a lack of research on a large number of excited states and degenerate problems for multi-dimensional complex eigenvalue problems. We use the multi-output neural network as multiple eigenstate ansatz functions, embed the boundary conditions into the ansatz functions, calculate the sum of squares of the equation residuals of each energy level on the spatial sampling point, calculate the orthogonal normalization error and energy uncertainty by Monte Carlo integral, minimize the weighted sum of these three items by gradient descent, and finally 
acquire the solution of multiple energy levels in a single training. At the same time, a method of obtaining multiple states from multiple training is given. Energy uncertainty has been added to the loss function to improve convergence accuracy. And we discussed the situations and approaches to degenerate levels in multi-dimensional problems. This method does not directly adopt the variational Monte Carlo (VMC) method, thus the sampling accuracy requirements are low, which can further save computing resources. And because the solution occurs at the zero point of each item, it is possible to solve the excited state without having to consider the weighting coefficients of each energy level. For simple boundary conditions, we generally use terms that multiply the solution function to meet the boundary conditions so that the solution function automatically satisfies the boundary conditions. For complex boundary conditions, points are generally taken at the boundary according to the idea of supervised learning so that the solution function meets the boundary conditions at these points. In order to verify the effectiveness of the algorithm, we first solved the problems of 1, 2, 3, 5, and 10 dimensions on the simple system of harmonic oscillators. For the one-, two-, and five-dimensional harmonic oscillator problems, our energy average absolute error is 0.0070\%, 0.0385\%, and 0.0306\%, respectively, and the reference \cite{li_neural-network-based_2021} energy average absolute error is 0.3850\%, 0.5927\%, and 1.6598\%, respectively. Especially for one-dimensional problems, we got this result after more than 30,000 iterations, and the reference\cite{li_neural-network-based_2021} got this result after more than 70,000 iterations. Fewer iterations shows that our method can achieve higher accuracy convergence with less resource consumption. Subsequently, we generalized the method to the hydrogen atom problem in Cartesian coordinate system, and successfully obtained the solution of the lowest five energy levels, which indicates that our method has the ability to solve singularity-containing potentials, and further shows that our algorithm has the potential to be extended to calculate the multi-excited states of the Schr\"{o}dinger equation for multi-electron atomic molecules.

\section{Results}\label{sec:results}

We first verify the feasibility of our algorithm in a simple harmonic oscillator system and determine the numerical ranges of some key hyperparameters. Then it is extended to the actual system like hydrogen atom and the wave function and energy calculation of the first few energy levels are completed. All the problems are solved directly by using Cartesian coordinate system. According to the analysis in Section \ref{sec:method}, for any potential field, we need to first determine the form of potential function and boundary conditions, then set reasonable boundary terms according to the potential field and boundary conditions, and design reasonable eigenvalue terms according to the possible forms of eigenvalues to speed up the convergence rate. It is also necessary to select a suitable sampling scheme according to the characteristics of the potential field. These steps are also illustrated in the following examples.

\subsection{Multi-dimensional uncoupled harmonic oscillator potential}\label{subsec:Multi-dimensional uncoupled harmonic oscillator potential}

We first calculate the classical problem of harmonic oscillator. To make simpler, let $m=\bar{h}=\omega=1$, the harmonic oscillator potential function is $V(x)={1\over 2}\sum_{i=1}^D x_i^2$, the wave function is $\Psi(\vec{R})$. It is easy to know that the term satisfying the boundary condition is $\mathcal{B}_n(x)=exp(-{1\over 2}\sum_i^D{\nu_{ni}x_i^2})$. And the real wave function of $D$-dimensional harmonic oscillator is  $\psi_n=\prod _{d=1}^D\psi_{ni}(x_d)$, the real energy is $E_n=\sum_{d=1}^Dn+{D\over 2}$, where $\psi_{nd}=N_{nd}H_{nd}(x_d)exp(-{1\over 2}x_d^2)$ is the real wave function of the energy level of the $d$th dimension,  is Hermitian polynomial, $N_{nd}$ is the normalization coefficient\cite{atakishiev_difference_1990,griffiths_schroeter_2018}. We calculate the wave functions and energy levels of 1, 2, 3, 5, and 10 dimension, respectively, and give the relationship between the key parameters and the convergence error.

The spatial sampling of this algorithm uses Gaussian sampling, and the density function is $P(X)=exp(-\sum_{d=1}^D({x_d\over\mu})^2)$, thus the weights of each sampling point are $\omega(X_s)={1\over P(X_s)}$. Through the change of each loss function with the number of iterations, we know what happens in the iterative convergence process, which reflects the performance characteristics of the solver and the reliability of the solution. The key parameter settings are shown in Table \ref{table.HO.parameter}. In pre-training, we generally adopted larger $lr$ and smaller $S$ to facilitate the fast convergence of the model, and larger sums to ensure that the wave functions of different energy levels can be obtained. During the transfer training, we reduced $lr^{2}$, $\alpha_2^{2}$ and $\alpha_3^{2}$, improved $S^{(2)}$, reduced $\alpha_2^{2}$ and $\alpha_3^{2}$ in order to improve the stability and accuracy of the convergence of each energy level in the pre-training.

\begin{figure}[H]
  \centering
    \includegraphics[scale=0.7]{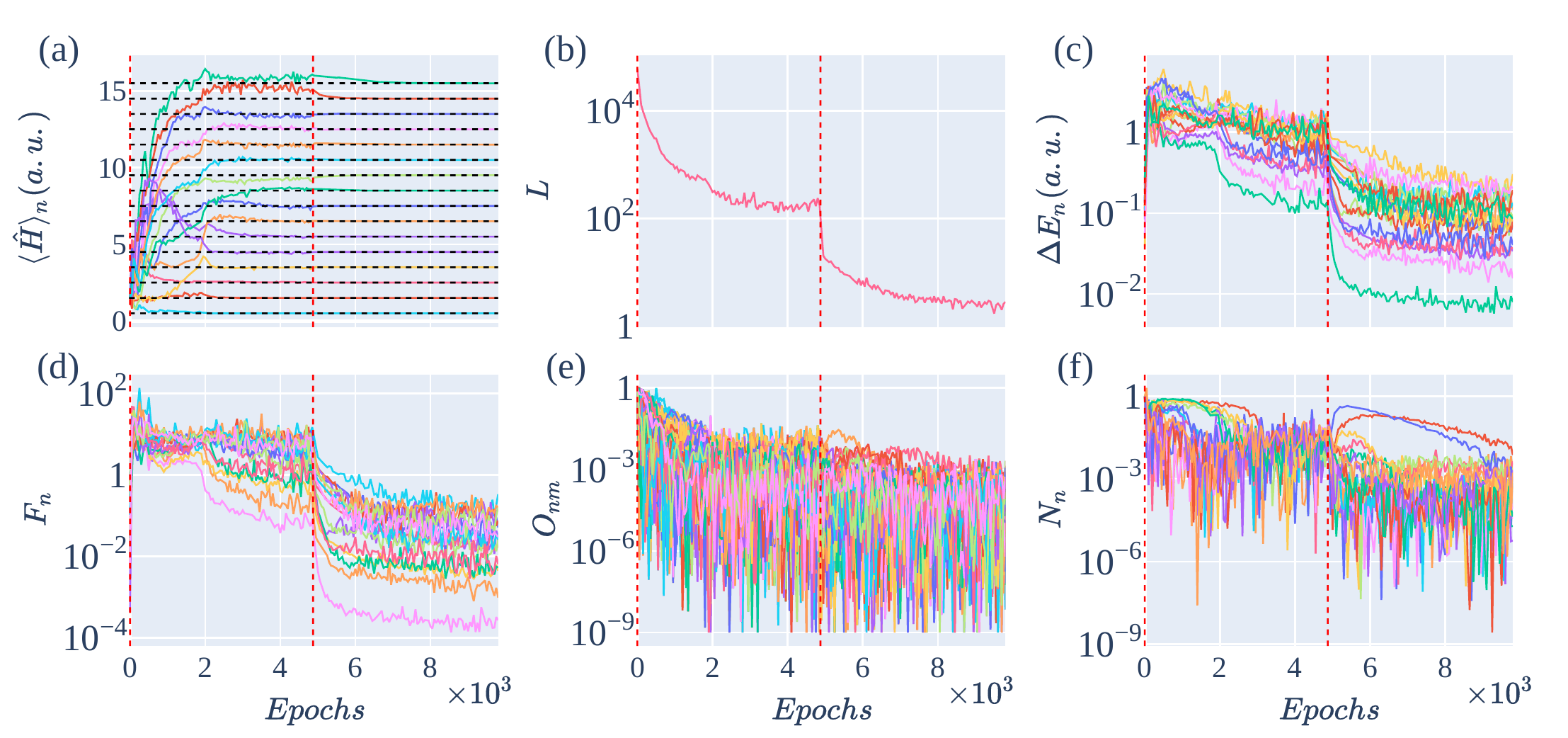}
  \caption{Convergence curve of one-dimensional harmonic oscillator, (a) is the average energy convergence curve, the black dotted line is the real energy level, (b) is the total loss function convergence curve, (c) is the energy uncertainty convergence curve, (d) is the residual equation convergence curve, (e) orthogonal condition convergence curve, (f) is the normalization condition convergence curve. The red vertical lines in figures represent the iteration starting points.}
  \label{fig.cerve.HO.1d}
\end{figure}

\begin{figure}[H]
  \centering
    \includegraphics[scale=0.7]{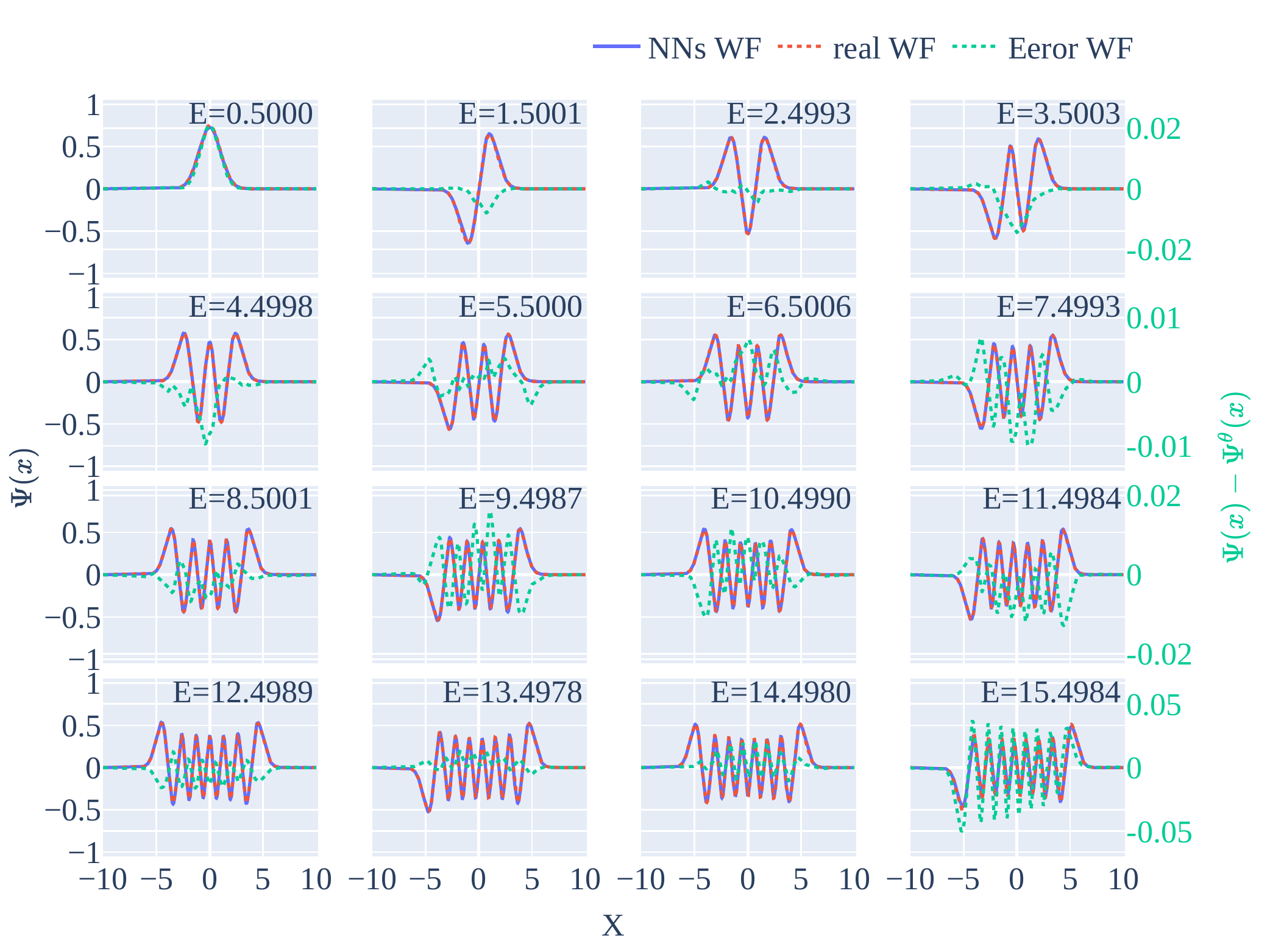}
  \caption{The wave function of all the first 16 energy levels of a one-dimensional harmonic oscillator and its error representation. The green dotted line in the figure shows the wave function error $\psi_n(x)-\psi_n^{\theta}(x)$, and its ordinate is shown on the right. The solid blue line is the neural network solution $\psi_n^{\theta}(x)$, the red dashed line is the analytical solution $\psi_n(x)$, and the ordinate axis is shown on the left. $E$ in each subplot is the corresponding energy of the wave function, arranged from the smallest to the largest.}
  \label{fig.psi.HO.1d}
\end{figure}

In the calculation of the one-dimensional harmonic oscillator, the wave function and energy convergence curves of 16 energy levels are given, as shown in Figures \ref{fig.cerve.HO.1d} and \ref{fig.psi.HO.1d}. From Figure \ref{fig.cerve.HO.1d}(a), the convergence process will initially undergo a disorderly convergence, and then $\psi_n^{\theta}(X_s)$ will continue to appear at each energy level, and the overall trend will gradually transition from low energy level to high energy level. Since it is not specified which energy level the output wave function corresponds to, there is also a situation of energy level exchange, which is similar to the situation shown in the reference\cite{li_neural-network-based_2021}. Note that our convergence does not stipulate that it must converge to a minimum of $N$ energy levels, but in the one-dimensional case, it still occupies the lowest energy level as possible. This is consistent with the assumption we introduced in Section \ref{subsec:theory}. Finally, the $E_n^{\theta}$ are sorted to determine the corresponding analytic wave function. At the same time, it is also necessary to determine the subspace of each energy level in the case of degeneracy, the specific method is shown in the Supplementary Information. $lr$ and $\alpha$ jointly control the energy level transition speed, while $lr$, $\alpha_1$, and $\alpha_2$ control convergence stability and convergence accuracy. Since our method uses Monte Carlo integrals, it is clear that the number of sampling points and sampling methods directly affect the convergence accuracy. Figure \ref{fig.cerve.HO.1d}(b) shows that there is a mutation in the total loss function $L$ during transfer training, which is caused by the decrease of the weight coefficient of the loss term, but it can be seen that the attenuation trend of $L$ also has a mutation, which is because after reducing the loss weight coefficient of the orthogonal normalization condition, Figure \ref{fig.cerve.HO.1d}(c) and (d) $F$ and $\Delta E$ become the main loss terms, which are rapidly reduced, so that the accuracy of the wave function is rapidly improved, and at the same time, the orthogonal condition Figure \ref{fig.cerve.HO.1d}(e) is naturally easy to meet.

Figure \ref{fig.psi.HO.1d} shows that the maximum error of the wave function is within 2\%, and the larger error occurs in the place where the gradient is large, which is consistent with the point proposed in the literature \cite{F_Principle,xuDeepFrequencyPrinciple2020,xuTrainingBehaviorDeep2019}. In addition, as the energy level increases, the error is also increasing, which is also the reason why it is more difficult to fit high frequencies.We further examine all the top 16 energy level convergence results for the five cases. From Table \ref{table.HO.energy}, it can be seen that there is no degeneracy in the one-dimensional case, $n$=0, $n$=1--2, $n$=3--5, $n$=6--9, $n$=10--14, $n$=15--16, respectively, the 0th, 1st, 2nd, 3rd, 4th, 5th, 6th energy levels, respectively, degeneracy degrees are 1, 2, 3, 4, 5, 2. In three-dimensional cases, $n$=0, $n$=1--3, $n$=3--8, $n$=9--15 are the 0th, 1st, 2nd, 3rd energy levels, respectively, and the degeneracy degree is 1, 3, 5, 7. In the case of 5 dimensions, $n$=0, $n$=1--5, $n$=6--15 are the 0th, 1th, and 2nd energy levels, respectively, and the degeneracy degree is 1, 5, 10. In the case of 10 dimensions, $n$=0, $n$=1--10, $n$=11--15 are the 0th, 1th, 2nd energy levels, respectively, and the degeneracy degree is 1, 10, 5. Since the analytical solution of the second energy level in 3D is a sixfold degenerate, the numerical solution has a state that converges to a higher energy level. The other cases are all satisfied with filling from low energy level to high energy level, which shows the correctness of the two assumptions introduced in Section \ref{sec:method}.

\subsection{Three-dimensional hydrogen-like atomic potential}\label{subsec:Three-dimensional hydrogen-like atomic potential}

The calculation of hydrogen-like atomic potential lays a certain foundation for the calculation of atomic molecular structure. In atomic units, hydrogen-like atomic potentials have $V(r)=-{Z\over r}$, where $r=\sqrt{x^2+y^2+z^2}$ is the radius and $Z$ is the nuclear charge number. The boundary condition item attached to the network is $\mathcal{B}_n^{\theta}=exp(-{r\over \nu})$, where $\nu$ is the parameter to be trained. The real wave function has $\psi_{nlm}(r)=N_{nl}({Zr\over na})^lexp(-{Zr\over na})F(-n+l+1,2l+2,2{Zr\over na})$ with an energy level of $E_n=-{Z^2\over n^2}{e^2\over 2a}$. 
Where $a={\hbar \over {m_e e^2}}$, $F(-n+l+1,2l+2,2{Zr\over na})$ are confluent hypergeometric functions and $N_{nl}={2Z^{3/2}\over{a^{3/2}n^2(2l+1)!}}\sqrt{{(n+1)!}\over{(n-l-1)!}}$ are normalization coefficients. We calculate the first 15 energy levels of H, He$^{+1}$, Li$^{+2}$, Be$^{+3}$, B$^{+4}$ and their wave functions, and give key parameters and error analysis.

\begin{figure}[H]
  \centering
  \subfigure[]{
    \label{fig:subfig:C2.3d} 
    \includegraphics[scale=0.6]{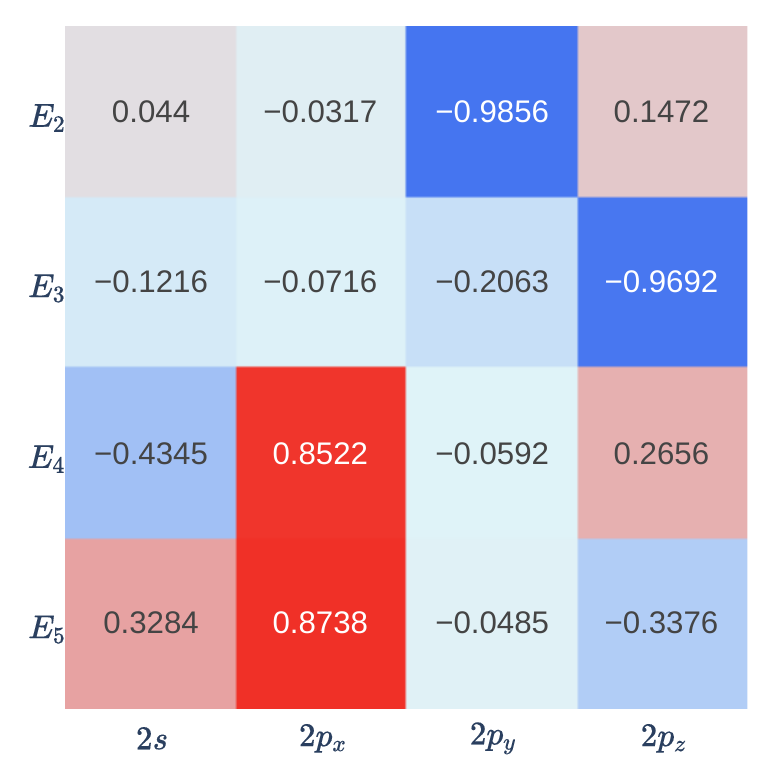}}
  \subfigure[]{
      \label{fig:subfig:C3.3d} 
      \includegraphics[scale=0.5]{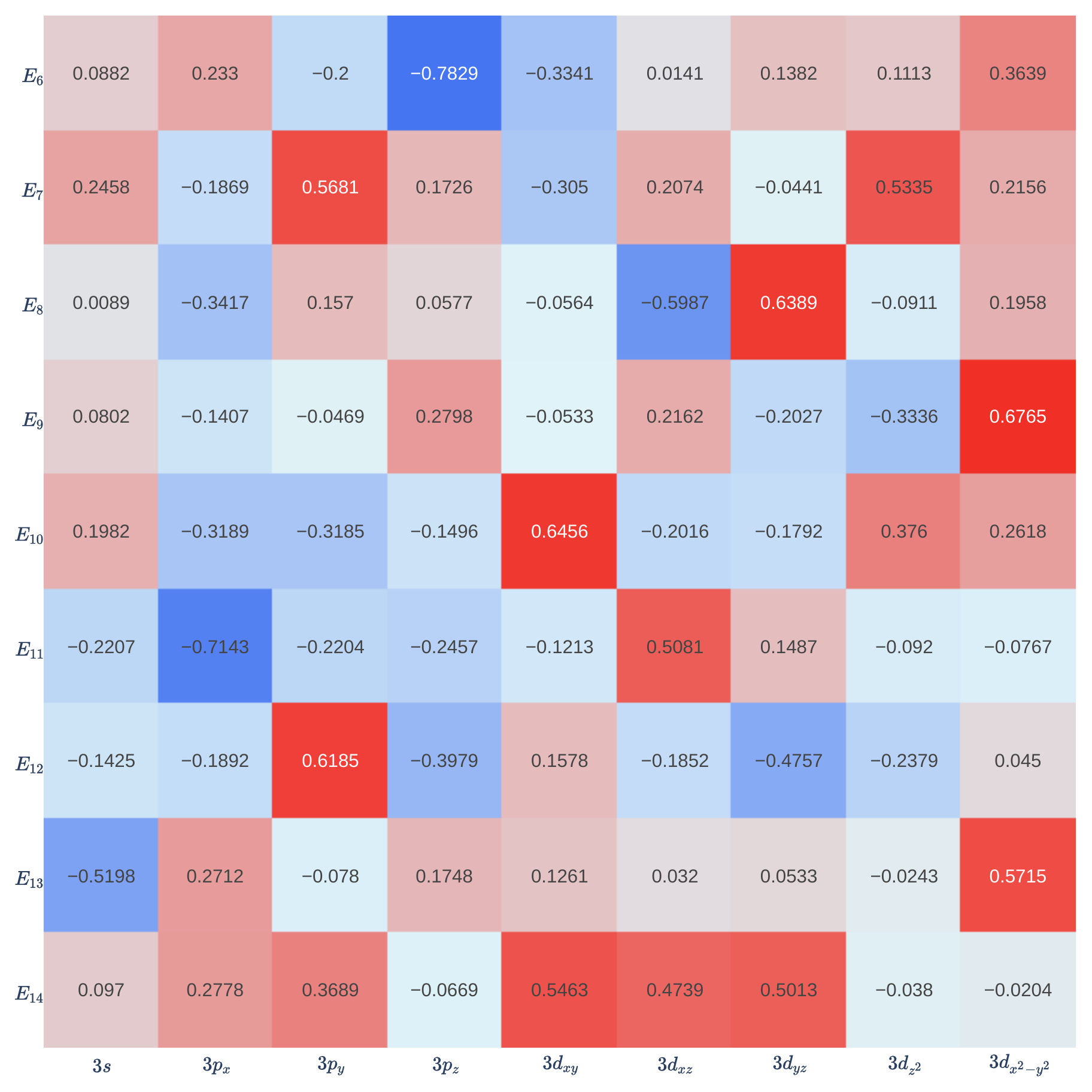}}

  \subfigure[]{
        \label{fig:subfig:C4.3d} 
        \includegraphics[scale=0.8]{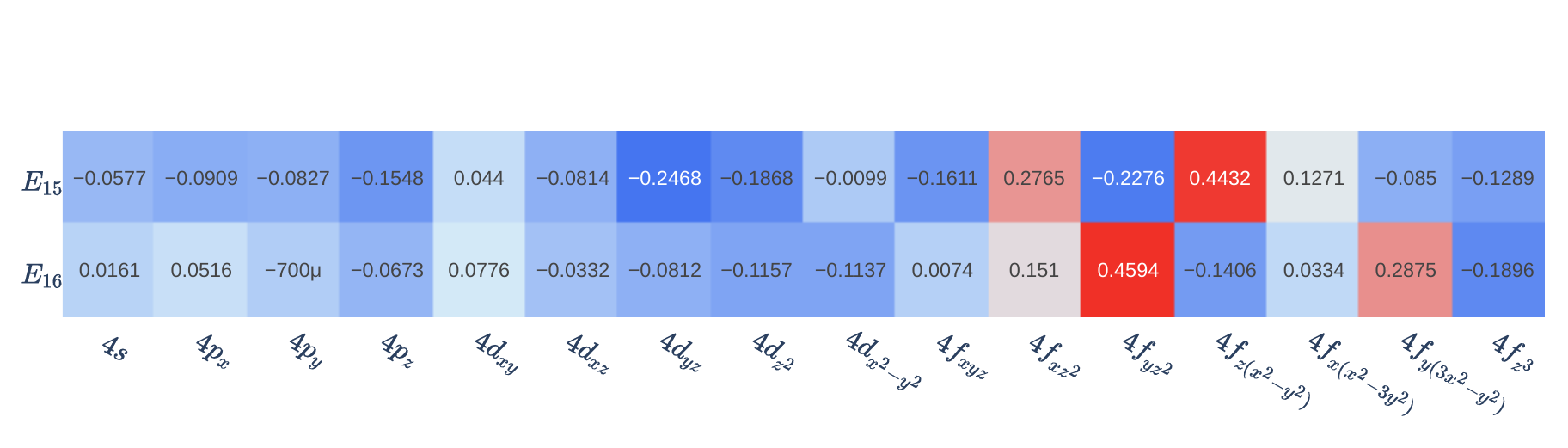}}
  \caption{(a) $n = 2$ energy level superposition coefficient, which is quadruple degeneration; (b) $n = 3$ energy level superposition coefficient, which is ninefold decentralization; (c) $n = 4$ energy level superposition coefficient, only the highest two energy levels.}
  \label{fig.C.Be.3d}
\end{figure}

We noted that for hydrogen-like atomic potentials, the zero point is a singularity, and in order to avoid model divergence, need truncation during the sampling, that is, only coordinates greater than \num{1d-3} can be taken. Referring to Section \ref{subsec:Multi-dimensional uncoupled harmonic oscillator potential}, we still use Gaussian sampling. Compared with the harmonic oscillator potential, the energy eigenvalue of the hydrogen-like atomic potential increases by the reciprocal of the square of the principal quantum number, which is inevitably difficult to converge in the highly excited state according to the theory of Section \ref{subsec:theory}. This is confirmed in Figure \ref{fig.C.Be.3d}. We only used simple Gaussian sampling, and only used thousands of points during pre-training to complete several wave function calculations of hydrogen-like atomic systems in three-dimensional space. The specific training parameters are shown in the Supplementary Information Table S1.

\begin{figure}[H]
  \centering
    \includegraphics[scale=0.7]{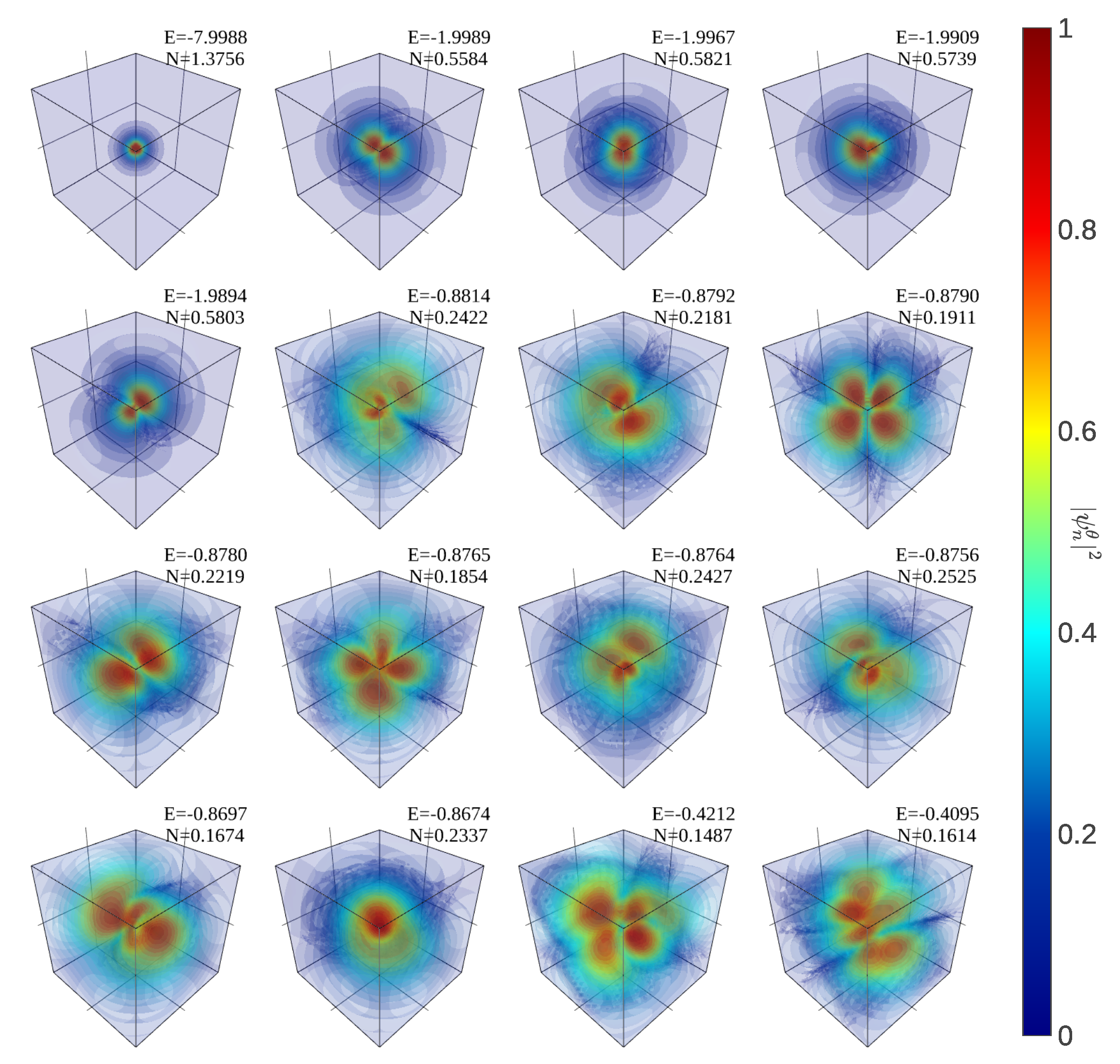}
  \caption{Be$^{+3}$ electron density outside the nucleus $|\psi_n^{\theta}|^2$ neural network numerical solution, the visualization area $x, y, z$ are from $-5$ to $+5$ (in atomic units) cube, $N$ is the probability density scaling coefficient in the figure, $E$ is the corresponding energy average of the wave function (in atomic units).}
  \label{fig.psi.Be.3d}
\end{figure}

We give the wave function for the first 16 energy levels of Be$^{+3}$, as shown in Figure \ref{fig.psi.Be.3d}. Then we further give the superposition coefficients of the four energy levels of Be$^{+3}$ ($n=2$), the nine energy levels of $n=3$, and the two energy levels (2) of $n=4$, as shown in Figure \ref{fig.C.Be.3d}. It is clear that with the exception of a very few states governed by atomic orbitals, most of the states are random combinations of atomic orbitals.

From Figure \ref{fig.psi.Be.3d}, the electron density structure of S, P, D, and F can be clearly observed, and the superposition characteristics of their interiors. The superposition coefficients of each degenerate level are further given, as shown in Figure \ref{fig.psi.Be.3d}, which are the four energy levels of $n=2$ of Be$^{+3}$, the 9 energy levels of $n=3$, and the superposition coefficient of two energy levels of $n=4$. Taking $E_3$ and $E_4$ as examples, we observe from Figure \ref{fig.psi.Be.3d} that their symmetry relative to the $p$ orbital is missing, because they are mainly superimposed by $2p_x$ and $2s$ orbitals. Without losing the generality, almost all degenerate wave functions are superimposed by the atomic orbitals of their corresponding energy levels, which is caused by the fact that this method does not take into account any functional symmetry. It can be noted that almost every state has a major component and several secondary components. 

As shown in Table \ref{table.H.energy}, the first 16 energy levels of the H atom only converge to the first five energy levels, and the subsequent energy levels cannot converge normally. A 3-energy level of a particular B$^{+4}$ ion converges to the 4th energy level, because our method itself does not specify convergence to the lowest energy level, but relies on the frequency principle of the neural network itself to make our energy level converge to the lowest possible level. This is not strict. For wave function error analysis, we use the treatment of degenerate wave function in Section \ref{subsec:Multi-dimensional uncoupled harmonic oscillator potential}. However, the two systems of He$^{+1}$ and Li$^{+2}$ have some anomalies, i.e., the energy of the highest two energy levels of these two systems is closer to $n=6$ and $n=5$, respectively, but their wave functions are closer to $n=4$ energy levels. Especially for hydrogen atomic systems, its energy level $n>2$ no longer converge to the bound state, but its wave function still has the characteristics of the bound state. From Table \ref{table.H.energy}, indicates that except for H atom the comprehensive evaluation index of other hydrogen-like ions is getting better with $Z$ increasing.

\section{Discussion}\label{sec:discussion}

\begin{figure}[H]
  \centering
  \includegraphics[scale=0.6]{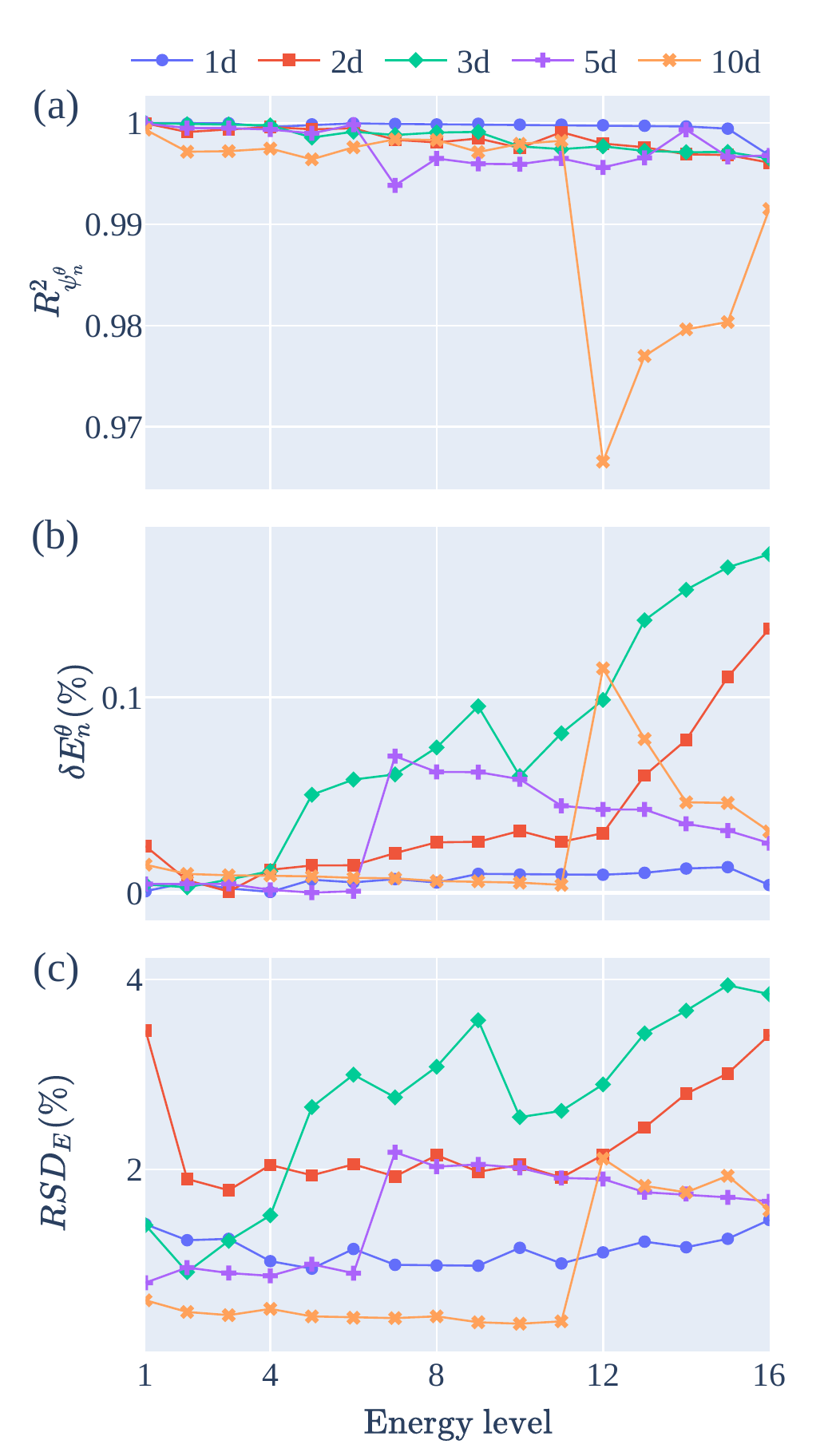}
  \caption{The wave function and energy level error of each energy level of harmonic oscillator potential, (a) $R^2$ in the figure represents the $R$-square of predicted wave function, which is used to characterize the wave function fitting performance; (b) $\delta E_n^{\theta}$ in the figure represents the absolute percentage error with the true energy level, which is used to characterize the accuracy of energy calculation; (c) RSD$_E$ represents the relative standard deviation of energy, which is used to characterize the comprehensive performance of the model.}
  \label{fig.Error}
\end{figure}

The comprehensive error analysis of wave function and energy eigenvalue for all the results of harmonic oscillators is shown in Figure \ref{fig.Error}. For harmonic oscillators, with the increase of energy level, except for the one-dimensional problem, (a) wave function error, (b) energy relative error and (c) energy uncertainty are significantly increased, and for uncoupled harmonic oscillators, it is known that its energy distribution is uniform, that is, the adjacent energy level difference is 1, thus the reason for this result is that the neural network is more difficult to fit a more complex function. In addition, it can be noted that when the amount of computation does not increase significantly, the convergence accuracy does not decrease as the dimension increases, which indicates that the method does have potential value for generalizing to high-dimensional problems. It is also noted that the relative energy uncertainty is correlated with the wave function error. (a) It is easy to notice that except for the $n>12$ state in 10 dimensions, the rest of the predicted wave function and analytical solution are greater than 0.99, which reflects the validity of our predicted wave function. The absolute percentage error of each energy level is less than 0.2\%, and the relative standard deviation of energy of each energy level is less than 4\%.

From Table \ref{table.HO.Error}, especially for the 1-, 2-, and 5-dimensional problems, our average absolute percentage error of energy is \num{6.9553d-5}, \num{3.8454d-4}, \num{3.0648d-4}, while the average absolute percentage error of energy for this problem is \num{3.8500d-3}, \num{5.9271d-3}, \num{1.6598d-2}. Interestingly, with convergence results of thousands of steps, our results are at least an order of magnitude more accurate. At the same time, good accuracy has been achieved in higher dimensional 10-dimensional problems, which can be regarded as almost unsolvable for finite element or finite difference algorithms, and there are still relatively few excited state studies for similar problems.

\section{Methods}\label{sec:method}

\subsection{Theory}\label{subsec:theory}

For any given SSE $\hat{H}\psi(\vec{R})=E\psi(\vec{R})$, where $\psi(\vec{R})$ is the required solution, $\hat{H}=\hat{T}+\hat{V}$ is the system Hamiltonian operator, and $E$ is the energy eigenvalue. There are $\vec{R}=(x_1,x_2,...,x_D)$ under Cartesian coordinates. The particle with a mass $m$ kinetic energy operator is $\hat{T}=-{1\over2}\sum_i^D{\partial^2\over\partial{x_i^2}}$, and $\hat{V}=V(\vec{R})$ is potential energy. Our problem lies in solving the first $N$ energy levels of this eigenvalue problem $E_0\leq E_1\leq E_2...\leq E_{N-1}$ and their corresponding wave functions $\psi_0(\vec{R}),\psi_1(\vec{R}),\psi_2(\vec{R}),...,\psi_{N-1}(\vec{R})$.

We design a $D$-input $N$-output neural network $\psi^\theta(\vec{R})$ with parameters $\theta$ to represent $\psi(\vec{R})$. The input layer corresponds to $x_1,x_2,...,x_D$, and the output layer corresponds to $\psi_0(\vec{R}),\psi_1(\vec{R}),\psi_2(\vec{R}),...,\psi_{N-1}(\vec{R})$. When we determine the parameters $\theta$, our wave function $\psi^\theta(\vec{R})$ is uniquely determined, and $E_n^\theta$ can also be uniquely determined.

\begin{figure}[ht]
  \centering
  \includegraphics[scale=0.6]{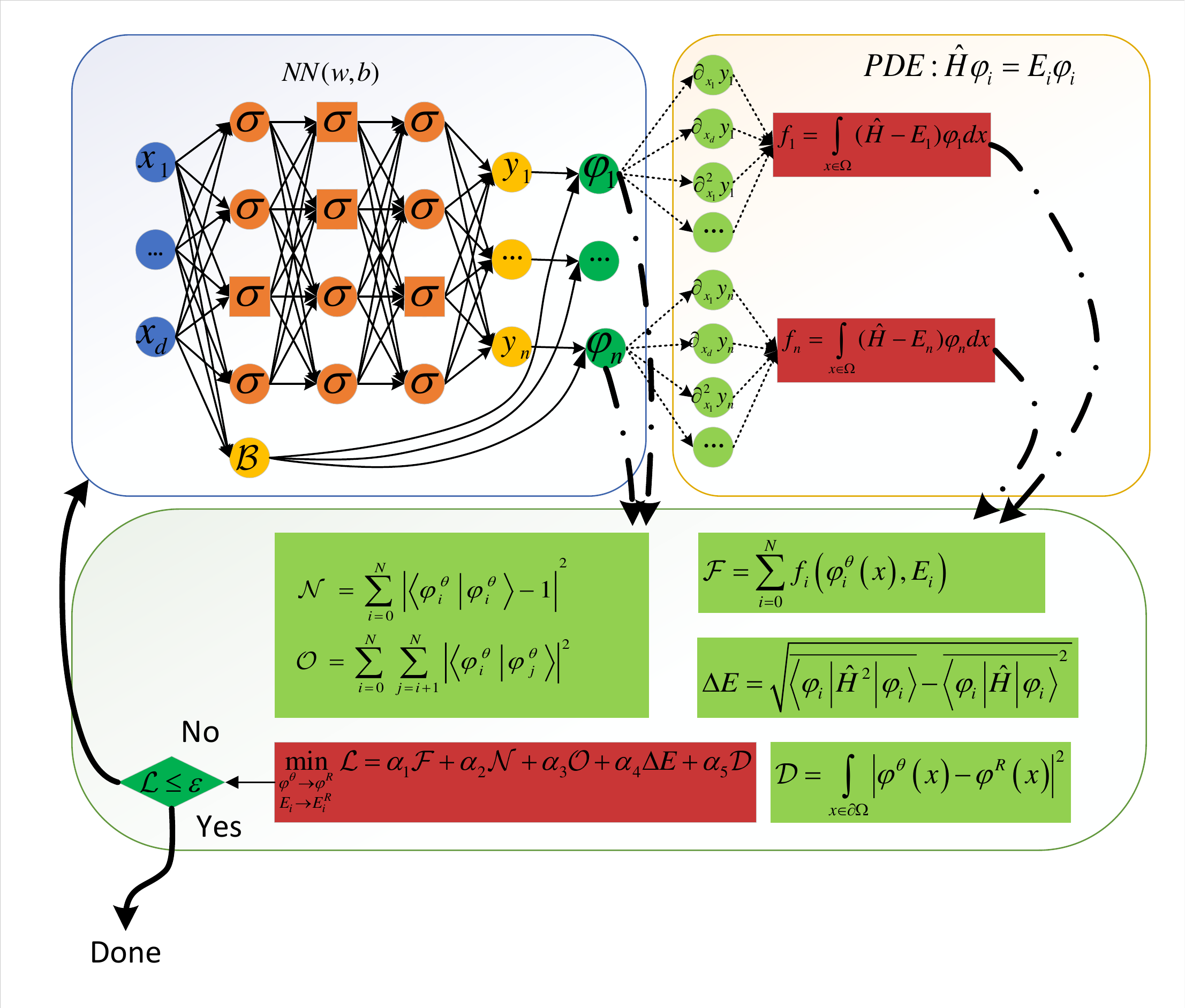}\\
  \caption{EPINNs principle, where $\Omega$ defines the field, $\mathcal{F}$ is the equation residuals, $\Omega_d$ is the known data coordinate domain of the equation solution, $\mathcal{D}$ is a data-driven term, generally containing boundary conditions, where $\psi_d(\vec{R})$ is the existing supervisory data, $\Delta E$ is the energy eigenvalue uncertainty, and $\mathcal{N}$ for the normalized conditional loss of the wave function, $\mathcal{O}$ is the orthogonal conditional loss. }
  \label{fig.epinns}
  \end{figure}

According to the standard process of neural network solving, we need to transform the solution of the SSE into an optimization problem, and then solve the eigenvalue problem by solving the optimization problem, and the algorithm flow is shown in Figure \ref{fig.epinns}.

We construct the equation residual loss term $\mathcal{F}=\sum_{n=1}^N{\langle \psi_n^\theta|(\hat{H}-E_n^\theta)^2|\psi_n^\theta\rangle}$. To make $\Psi_n^\theta(\vec{R})$ and $E_n^\theta$ satisfy the equation; different solutions must be orthogonal to each other, we construct orthogonal loss terms $\mathcal{O}=\sum_{n=1}^N\sum_{m=n+1}^N\| {\langle \psi_n^\theta|\psi_m^\theta\rangle}\Vert $; the same solution is normalized, we construct normalized loss terms $\mathcal{N}=\sum_{n=1}^N\| {\langle \psi_n^\theta|\psi_n^\theta\rangle}-1\Vert $; to satisfy appropriate boundary conditions with the solution, we construct supervised learning loss terms $\mathcal{D}={1\over N}\sum_{n=1}^N(\int_{x \in \partial\Omega}\| \psi_n^\theta-\psi_n\Vert dx+\int_{x \in \partial\Omega}\| \partial_R\psi_n^\theta-\partial_R\psi_n\Vert dx)$; finally, in order to improve the convergence accuracy, we construct the energy uncertainty $\Delta E=\sum_{n=1}^N\sqrt{{\langle\psi_n^\theta|\hat{H}^2|\psi_n^\theta\rangle}-{\langle\psi_n^\theta|\hat{H}|\psi_n^\theta\rangle}^2}$; in summary, the solution we need can be obtained by adjusting $\theta$ to minimize the following loss function $\mathcal{L}$.
$\mathcal{L}=\alpha_1 \mathcal{F}+\alpha_2 \Delta E+\alpha_3 \mathcal{N}+\alpha_4 \mathcal{O}+\alpha_5 \mathcal{D}.$

In our problem, in order to accelerate the convergence rate, we embed the terms satisfying the boundary conditions into the ansatz function. Therefore, the supervised learning loss term is automatically constant 0. Then the final loss function is
$$\mathcal{L}=\alpha_1 \mathcal{F}+\alpha_2 \Delta E+\alpha_3 \mathcal{N}+\alpha_4 \mathcal{O}.$$

It can be proved that, when $\mathcal{L}=0$, $E_n^{\theta}$ and $\psi_n^{\theta}(\vec{R})$ can obtain the exact solution.

Let $\psi_n^{\theta}(\vec{R})=\psi_n(\vec{R})+\sum_{i\neq n}c_i\psi_i(\vec{R})$, $E_n^{\theta}=E_n+\delta E_n$, where $\psi_n(\vec{R})$ is the real normalized eigen wave function of the energy level $n$, $E_n$ the real energy level. Obviously $c_i$ is a small amount. And then we have:

The equation residual of $n$ energy level is
$$\mathcal{F}={\langle \psi_n^\theta|(\hat{H}-E_n^\theta)^2|\psi_n^\theta\rangle}=\sum_{i\neq n}^N{c_i^2(E_i-E_n-\delta E_n)^2}+\delta E_n^2.$$

The energy uncertainty of $n$ energy level is
$$\Delta E=\sqrt{{\langle\psi_n^\theta|\hat{H}^2|\psi_n^\theta\rangle}-{\langle\psi_n^\theta|\hat{H}|\psi_n^\theta\rangle}^2}=\sqrt{\sum_{i\neq n}^Nc_i^2(2E_n-E_i)E_i+|\sum_{i\neq n}^N{c_i^2E_i|^2}}.$$

The normalization error of the energy level $n$ is
$$\mathcal{N}=\sum_{n=1}^N\| {\langle \psi_n^\theta|\psi_n^\theta\rangle}-1\Vert =\|\sum_{i\neq n}c_i^2\Vert.$$

The orthogonal error of the energy level $n,m$ is
$$\mathcal{O}=\sum_{n=1}^N\sum_{m=n+1}^N\| {\langle \psi_n^\theta|\psi_m^\theta\rangle}\Vert =\|c_n+c_m+\sum_{i\neq n,m}c_i^2\Vert.$$

It is easy to know from the above results that if and only if all sum $c_i=0$ and $\delta E_n=0$, our loss function $\mathcal{L}=0$. And it is easy to get that although the final convergence result has nothing to do with the properties of energy levels, the convergence process is affected by the order of magnitude of the difference between the desired energy levels $E_i-E_n$ and the values of other energy levels $E_i^2$ from $\Delta E_n$ and $F_n$. This property will have a significant impact on the convergence when solving the hydrogen atom system.

Compared with the convergence result of the orthogonal normalization condition introduced when solving the multiple excited states, the direct application of the variational method $\langle\psi_n^{\theta}|\hat{H}|\psi_n^{\theta}\rangle=E_n+\sum_{i\neq n}^Nc_i^2E_i$ is easy to see the influence of a large number $E_n$, which will make the convergence of each loss unbalanced, and ultimately directly affect the convergence result.

\subsection{Numerical methods}\label{subsec:method}

We defined the ansatz wave function $\Psi_n^{\theta}(\vec{R})=N_n^{\theta}(\vec{R})\cdot \mathcal{B}_n^{\theta}(\vec{R})$ based on back propagation neural network (BPNN) as shown in Figure \ref{fig.epinns}, where $N_n^{\theta}(\vec{R})$ is an output of a multi-output BPNN, whose main function is to map the input layer nonlinearly, and to approximate the details of the $N$-level wave function by the universal approximation theorem of the neural network. $\mathcal{B}_n^{\theta}(\vec{R})$ is a manually set term satisfying the boundary conditions. The main function of this term is to force the ansatz wave function to satisfy the boundary conditions. In order to ensure the smoothness and symmetry of the solution, we take $\tanh(x)$ as the activation function\cite{Jarrett2009WhatIT}. The fully connected neural network of layers (excluding input layer) with each node of hidden layer $Nh$ is used as $N_n^{\theta}(\vec{R})$. Hidden layer nodes have
$$h_i^{k+1}=\tanh(\sum_{j=1}^{Nh}w_{ij}^kh_j^k+b_i^k).$$

In order to make the solution to all real numbers, the activation function of the output layer selects linear function, which is
$$N_n^{\theta}=\sum_{j=1}^{Nh}w_{ij}^{Nl}h_j^{Nl}+b_n^{Nl},$$
where $w_{ij}^{k}$ is the weight of the $k$ layer, and $b_i^k$ is the bias term of the $k$ layer. Note that for different $V(\vec{R})$ and boundary conditions, we need to set different $\mathcal{B}_n^{\theta}(\vec{R})$. More specific details are discussed in Section \ref{sec:results}.

Due to the complexity of the proposed wave function itself and the requirement of stochastic gradient descent, automatic differential technology \cite{paszkePyTorchImperativeStyle} and Monte Carlo integral \cite{GEWEKE1996731} should be adopted in the calculation of various loss functions. It is noted that this method is not sensitive to the integration accuracy, and in order to improve the sampling efficiency, we directly use the general random sampling method instead of Markov chain Monte Carlo sampling. There are two kinds of sampling, one is Gaussian sampling under infinite boundary conditions, and the other is uniform sampling under finite boundary conditions. Relying on the torch library to quickly generate the points needed directly on video memory, rather than generating data in memory and then moving it to video memory, reduces unnecessary overhead.

For each loss function in Section \ref{subsec:theory}, we need to carry out discrete calculation. Let us say the sampling points have $X_s$, and the weight of each point in the whole space has $\omega(X_s)$, thus we have:

Discrete form of equation residuals $F_n={1\over S}\sum_{s=1}^S{\omega(X_s)|\hat{H}\psi_n^{\theta}(X_s)-E_n^{\theta}\psi_n^{\theta}(X_s)|}$.

Discrete form of orthogonal conditional $O_{nm}={1\over S}\sum_{s=1}^S{\omega(X_s)|\psi_n^{\theta}(X_s)\psi_m^{\theta}(X_s)|^2}$.

Discrete forms of normalized conditional $N_{n}=|{1\over S}\sum_{s=1}^S{\omega(X_s)|\psi_n^{\theta}(X_s)|^2-1|^2}$.

Discrete form of energy uncertainty $\Delta E_n=\sqrt{{\langle\hat{H^2}\rangle}_n-{\langle\hat{H}\rangle}_n^2}$,\\ where ${\langle\hat{H^2}\rangle}_n=\bigg||\hat{H}\rangle_n\bigg|^2={1\over S}\sum_{s=1}^S{\omega(X_s)\big|\hat{H}\psi_n^{\theta}(X_s)\big|^2}$, ${\langle\hat{H}\rangle}_n={1\over S}\sum_{s=1}^S{\omega(X_s)\psi_n^{\theta}(X_s)\hat{H}\psi_n^{\theta}(X_s)}$.

Therefore, the discrete form of the total loss function is,
$$L=\alpha_1 \sum_n F_n+\alpha_2 \sum_n\Delta E_n+\alpha_3 \sum_n N_n+\alpha_4 \sum_{nm} O_{nm}.$$

The iterative solution process is similar to the conventional neural network training process, i.e., it mainly includes spatial random sampling, forward propagation, loss function calculation, key index storage, convergence judgment, back propagation.

In this case, the convergence criterion is stable for a period of time, which is that the energy deviation of the iteration is less than the threshold $\sqrt{{1\over N}\sum_{n=1}^N{1\over{T_e}}\sum_{t=T}^{T-T_e}(E_n^t-\overline{E_n})^2}\leq \varepsilon$.

In this paper, Xavier \emph{et al.}'s work\cite{Xavier} is used to initialize neural network parameters. As a whole, the training process is divided into two steps. The first step is to apply large coefficients to the orthogonal normalization conditions and the residual of the equation, and the number of sampling points can be appropriately reduced, so that each output can converge to each mutually orthogonal energy level. Then, by reducing these three coefficients, increasing the uncertainty coefficient of the eigenvalue and the number of sampling points, the final convergence accuracy can be improved.

It is well known that the solution space of neural network is a high-dimensional hypersurface. The solution space of the VMC method has no zero point, but its optimal solution is unique. The weight of each loss function needs to be set carefully, because it will have a great influence on the results. The solution space of this method is a hypersurface with infinite zeros, and the optimal solution is these zeros. Each weight coefficient only affects the convergence process, but not the convergence result directly. Our initialization method will make the solution fall near the ground state solution, so all the states will converge to the ground state at the beginning, and then the residual term and the normalized condition term of the equation will tend to zero, while the orthogonal condition term accounts for the main component of the loss function. Then, due to the existence of the orthogonal condition, each solution will be excited to a higher energy level until it occupies a lower $N$ energy levels in turn. This is the reason for the two-step method.

\section*{Data availability}
All data that support the findings of this study are included in the paper are available from the corresponding author upon request.

\section*{Code availability}
The codes used to generate results are available from the corresponding author on request.

\section*{Supplementary information}
The online version contains supplementary material available at 

\section*{Acknowledgements}
J.L. thanks Fan Wang and Xi He for helpful discussions. C.Y. was supported by the Fundamental Research Funds for the Central Universities (Grant No. 10822041A2038).

\section*{Author contributions}
C.Y. and G.J. conceived the project. J.L. designed the neural network, implemented the algorithms and wrote the code. J.L. fulfilled the calculations and interpreted data. J.L., X.D. and C.Y. discussed the results. J.L. and X.D. wrote the manuscript. All authors commented on the manuscript.

\section*{Competing interests}
The authors declare no competing interests.



\begin{table}[!ht]\centering
  \caption{Key parameters of the multi-dimensional harmonic oscillator, where $lr$ represents the learning rate of the pre-training, $La$ represents the depth of the network (except the input layer), represents the number of nodes of the hidden layer, $Nh$ represents the number of energy levels solved in a single time, $Nl$ represents the weight of each loss function of the pre-training, $S$ represents the number of sampling points of the pre-training, $lr^{(2)}$ represents the transfer learning rate, $\alpha^{(2)}$ represents the weight of each loss function of the transfer learning , $S^{(2)}$ represents the number of migration learning sampling points, $\mu$ represents the Gaussian sampling half-width.}
  \label{table.HO.parameter}
  \begin{tabular}{cccccc}
    \toprule
    & 1d & 2d & 3d & 5d& 10d\\
    \midrule
    $lr$&\num{5d-3}&\num{1d-3}&\num{5d-3}&\num{5d-3}&\num{5e-3}\\
    $La$&3&3&3&3&3\\
    $Nh$&150&150&150&150&150\\
    $Nl$&16&16&16&16&16\\
    $\alpha$&1,10,250,1&1,10&1,2,10,1&1,2,10,1&1,2,10,1\\
    $S$&2048&8192&2048&2048&2048\\
    $lr^{(2)}$&\num{1d-3}&\num{5d-4}&\num{5d-4}&\num{5d-4}&\num{5d-4}\\
    $\alpha^{(2)}$&1,1,1,1&1,1,1,1&1,1,1,1&1,1,1,1&1,1,1,1\\
    $S^{(2)}$&8192&2048&8192&8192&8192\\
    $\mu$&4&3&2&2&2\\
    \bottomrule
    \end{tabular}
\end{table}

\begin{table}[!ht]\centering
  \caption{The neural network numerical solution of energy eigenvalues (in reduced units), the first column $n$ is the energy level ordinal number, from the ordering of energy. Columns 2 to 6 are the average energy of the 1, 2, 3, 5, and 10-dimensional harmonic oscillators, derived from the average energy calculated by random sampling of 300 transfer learning parameters, and the uncertainty of the 300th average energy in parentheses, which comes from the standard deviation of the energy of the last 300 iterations.}
  \label{table.HO.energy}
  \begin{tabular}{ccccccccccc}
    \toprule
    &\multicolumn{2}{c}{1d} & \multicolumn{2}{c}{2d} & \multicolumn{2}{c}{3d} & \multicolumn{2}{c}{5d}& \multicolumn{2}{c}{10d}\\
    &$E_n^{\theta}$&$E_n$&$E_n^{\theta}$&$E_n$&$E_n^{\theta}$&$E_n$&$E_n^{\theta}$&$E_n$&$E_n^{\theta}$&$E_n$\\
    \midrule
    0&  0.5000(5)   &0.5&1.0002(12)&1.0 &1.5001(6) &1.5 &2.5001(2) &2.5 &5.0007(9) &5.0\\
    1&  1.4999(6)   &1.5&1.9999(12)&1.9 &2.5001(6) &2.5 &3.4998(5) &3.5 &5.9994(9) &6.0\\
    2&  2.5001(11)  &2.5&2.0000(12)&2.0 &2.5002(7) &2.5 &3.4998(4) &3.5 &5.9995(9) &6.0\\
    3&  3.5000(10)  &3.5&3.0004(13)&3.0 &2.5003(9) &2.5 &3.4999(5) &3.5 &5.9995(11)&6.0\\
    4&  4.4997(13)  &4.5&3.0004(9) &3.0 &3.5018(11)&3.5 &3.5000(4) &3.5 &5.9995(9) &6.0\\
    5&  5.4997(13)  &5.5&3.0004(10)&3.0 &3.5020(14)&3.5 &3.5000(4) &3.5 &5.9995(7) &6.0\\
    6&  6.4995(15)  &6.5&4.0008(6) &4.0 &3.5021(16)&3.5 &4.4969(10)&4.5 &5.9996(7) &6.0\\
    7&  7.4996(16)  &7.5&4.0010(12)&4.0 &3.5026(14)&3.5 &4.4972(10)&4.5 &5.9996(8) &6.0\\
    8&  8.4992(17)  &8.5&4.0010(8) &4.0 &3.5033(22)&3.5 &4.4972(9) &4.5 &5.9996(12)&6.0\\
    9&  9.4991(14)  &9.5&4.0013(19)&4.0 &4.5027(14)&4.5 &4.4974(10)&4.5 &5.9997(7) &6.0\\
    10& 10.4990(18) &10.5&5.0013(18)&5.0 &4.5037(17)&4.5 &4.4980(5) &4.5 &5.9998(10)&6.0\\
    11& 11.4989(18) &11.5&5.0015(7) &5.0 &4.5044(19)&4.5 &4.4981(8) &4.5 &6.9920(22)&7.0\\
    12& 12.4987(27) &12.5&5.0030(17)&5.0 &4.5063(27)&4.5 &4.4981(6) &4.5 &6.9945(18)&7.0\\
    13& 13.4983(22) &13.5&5.0039(13)&5.0 &4.5070(17)&4.5 &4.4984(7) &4.5 &6.9968(30)&7.0\\
    14& 14.4981(21) &14.5&6.0066(24)&6.0 &4.5075(22)&4.5 &4.4986(5) &4.5 &6.9968(14)&7.0\\
    15& 15.4994(28) &15.5&6.0081(9) &6.0 &4.5078(25)&4.5 &4.4989(3) &4.5 &6.9978(14)&7.0\\
    \bottomrule
    \end{tabular}
\end{table}

\begin{table}[!ht]\centering
  \setlength{\tabcolsep}{4pt}
  \caption{The average energy of each energy level of the hydrogen-like atomic system (in atomic units), $n$ in the first column is the sequence number of the energy level, and columns 2--6 are the average energy of each energy level of H--B$^{+4}$, and the standard deviation of the average energy calculated by sampling according to the transfer learning parameters is 300 times in parentheses. Energy levels 0 in the table are all ground states, energy levels 1--4 are the first excited state, except H, B$^{+4}$ energy levels 5--13 are the second excited state, except H, B$^{+4}$ energy levels 14--15 are the third excited state, and one of the second excited states in B$^{+4}$ converges to the third excited state.}
  \label{table.H.energy}
  \begin{tabular}{ccccccccccc}
    \toprule
    &\multicolumn{2}{c}{H} & \multicolumn{2}{c}{He$^{+1}$} & \multicolumn{2}{c}{Li$^{+2}$} & \multicolumn{2}{c}{Be$^{+3}$}& \multicolumn{2}{c}{B$^{+4}$}\\

    &$|E_n^{\theta}|$&$|E_n|$&$|E_n^{\theta}|$&$|E_n|$&$|E_n^{\theta}|$&$|E_n|$&$|E_n^{\theta}|$&$|E_n|$&$|E_n^{\theta}|$&$|E_n|$\\
    \midrule
    0&0.4999(0)  &0.500  &1.9998(0)  &2.0 &4.4995(2) &4.500  &7.9988(44) &8.0 &12.4989(30)&12.50 \\  
    1&0.1225(24) &0.125  &0.4994(2)  &0.5 &1.1235(6) &1.125  &1.9989(61) &2.0 &3.1240(9)  &3.125 \\ 
    2&0.1219(17) &0.125  &0.4993(4)  &0.5 &1.1230(5) &1.125  &1.9967(48) &2.0 &3.1237(8)  &3.125 \\ 
    3&0.1216(49) &0.125  &0.4976(25) &0.5 &1.1215(43)&1.125  &1.9909(276)&2.0 &3.1235(11) &3.125 \\ 
    4&0.1190(56) &0.125  &0.4970(61) &0.5 &1.1199(72)&1.125  &1.9894(172)&2.0 &3.0883(65) &3.125 \\  
    5&&&0.2112(20)  &0.222 &0.4929(139) &0.5 &0.8814(53) &0.889 &1.3819(28) &1.389 \\ 
    6&&&0.2108(13)  &0.222 &0.4787(172) &0.5 &0.8792(57) &0.889 &1.3819(37) &1.389 \\ 
    7&&&0.2095(12)  &0.222 &0.4769(121) &0.5 &0.8790(46) &0.889 &1.3790(83) &1.389 \\ 
    8&&&0.2094(15)  &0.222 &0.4734(131) &0.5 &0.8780(41) &0.889 &1.3748(39) &1.389 \\ 
    9&&&0.2089(23)  &0.222 &0.4728(88)  &0.5 &0.8765(60) &0.889 &1.3693(51) &1.389 \\ 
    10&&&0.2071(25) &0.222 &0.4672(96)  &0.5 &0.8764(63) &0.889 &1.3645(74) &1.389 \\ 
    11&&&0.2069(21) &0.222 &0.4663(97)  &0.5 &0.8756(81) &0.889 &1.3627(49) &1.389 \\ 
    12&&&0.2067(24) &0.222 &0.4583(65)  &0.5 &0.8697(95) &0.889 &1.3509(153)&1.389 \\ 
    13&&&0.1837(73) &0.222 &0.4582(92)  &0.5 &0.8674(48) &0.889 &0.7363(249)&0.781 \\ 
    14&&&0.0921(82) &0.125 &0.1451(342) &0.281 &0.4212(323)&0.5 &0.7340(201)&0.781 \\ 
    15&&&0.0852(61) &0.125 &0.1340(308) &0.281 &0.4095(283)&0.5 &0.7253(213)&0.781 \\ 
    \bottomrule
    \end{tabular}
\end{table}

\begin{table}
\centering
  \caption{The comprehensive evaluation index of harmonic oscillator potential, where MAPE$_E$ is the average of the relative error of energy, MRSD$_E$ is the average of the standard deviation of energy, and $\overline{\delta\psi}$ is the average of the root mean square error of the wave function. $R_{\psi}^2$ is the $R$ side of a case's total. $Epoch$ is the number of pre-training iterations and $Epoch^{(2)}$ is the number of migration training iterations. $T$ is the pre-training iteration time, and $T^{(2)}$ is the migration training iteration time.}
  \label{table.HO.Error}
  \begin{tabular}{rrrrrr}
    \toprule
    & 1d & 2d & 3d & 5d& 10d\\
    \midrule

    $\overline{\delta\psi}$ &0.0136&0.0402&0.0424&0.0365&0.0685\\
    $R_{\psi}^2$ &0.9997&0.9984&0.9986&0.9980&0.9933\\
    MAPE$_E(\%)$ &\num{6.9553d-3}&\num{3.8454d-2}&\num{7.7344d-2}&\num{3.0648d-2}&\num{2.5258d-2}\\
    MRSD$_E(\%)$ &0.0115&0.0231&0.0269&0.0152&0.0089\\
    $Epoch$ &4873&2512&1238&1261&2312\\
    $Epoch^{(2)}$ &4957&1230&729&783&2663\\
    $T$ &1803&1343&859&1180&4977\\
    $T^{(2)}$ &1945&659&538&725&6121\\
    
    \bottomrule
    \end{tabular}
\end{table}

\end{document}